\documentclass[useAMS,usenatbib]{mn2e}
\usepackage{txfonts}
\usepackage[latin1]{inputenc}
\usepackage[T1]{fontenc}
\usepackage{ngerman}
\usepackage{mathptmx}
\usepackage[scaled]{helvet}
\usepackage{courier}
\usepackage{aecompl} 
\usepackage{longtable}  
\usepackage{rotating}
\usepackage{natbib}
\usepackage{graphicx}
\usepackage{graphics}
\usepackage{psfrag}
\usepackage{amssymb}
\usepackage{multicol}
\usepackage{multirow}
\usepackage{lscape}
\usepackage{color}
\usepackage{soul}
\bibpunct{(}{)}{;}{a}{}{,}
\def\Teff{\ensuremath{T_{\mathrm{eff}}}}

\def\logg{\ensuremath{\log g}}

\def\vmic{$\upsilon_{\mathrm{mic}}$}

\def\vsini{\ensuremath{{\upsilon}\sin i}}

\def\logt{\ensuremath{\log t}}

\title[]{HD~54272, a classical $\lambda$ Bootis star and $\gamma$ Doradus pulsator} 

\author[E. Paunzen et al.]{E. Paunzen,$^{1}$\thanks{epaunzen@physics.muni.cz}
			   M. Skarka,$^{1}$
                           D.L. Holdsworth,$^{2}$
                           B. Smalley,$^{2}$ and
			   R.G. West\,$^{3}$
			   \\
$^{1}$Department of Theoretical Physics and Astrophysics, Masaryk University,
Kotl\'a\v{r}sk\'a 2, 611\,37 Brno, Czech Republic\\
$^{2}$Astrophysics Group, Keele University, Staffordshire, ST5 5BG, UK\\
$^{3}$Department of Physics, University of Warwick, Coventry, CV4 7AL, UK}
\begin{document}

\date{}

\pagerange{\pageref{firstpage}--\pageref{lastpage}} \pubyear{2014}

\maketitle

\label{firstpage}

\begin{abstract}
We detect the second known $\lambda$ Bootis star (HD 54272) which exhibits $\gamma$ Doradus type pulsations.
The star was formerly misidentified as a RR Lyrae variable.
The $\lambda$ Bootis stars are a small group (only 2\%) of late B to early F-type, Population I
stars which show moderate to extreme (up to a factor 100) surface underabundances
of most Fe-peak elements and solar abundances of lighter elements (C, N, O, and S). 
The photometric data from the Wide Angle Search for Planets
(WASP) and All Sky Automated Survey (ASAS) projects
were analysed. They have an overlapping time base of 1566~d and 2545~d, respectively. 
Six statistically significant peaks were identified ($f_{1}=1.410116$\,d$^{-1}$, $f_{2}=1.283986$\,d$^{-1}$, 
$f_{3}=1.293210$\,d$^{-1}$, $f_{4}=1.536662$\,d$^{-1}$, $f_{5}=1.15722$\,d$^{-1}$ and $f_{6}=0.22657$\,d$^{-1}$). 
The spacing between $f_{1}$ and $f_{2}$, $f_{1}$ and $f_{4}$, $f_{5}$ and $f_{2}$ is almost identical. Since the 
daily aliasing is very strong, the interpretation of frequency spectra is somewhat ambiguous. 
From spectroscopic data, we deduce a high rotational velocity (250$\pm$25\,km\,s$^{-1}$)
and a metal deficiency of about $-$0.8 to $-$1.1\,dex compared to the Sun. A comparison with 
the similar star, HR 8799, results in analogous pulsational characteristics but widely different astrophysical
parameters. Since both are $\lambda$ Bootis type stars, the main mechanism of this phenomenon, selective accretion,
may severely influence $\gamma$ Doradus type pulsations.
\end{abstract}

\section{Introduction}\label{introduction}

The group of classical $\lambda$ Bootis stars comprises late B to early F-type, Population I
stars, with moderate to extreme (up to a factor of 100) surface underabundances
of most Fe-peak elements and solar abundances of lighter elements (C, N, O,
and S). They are rare, with a maximum of about 2\% of all objects in the relevant 
spectral domain, between the zero- and terminal-age main-sequence (ZAMS and TAMS), are found to be 
such objects (\citealt{Pau02b}). 

\citet{Mic86} suggested that the peculiar chemical 
abundances on the stellar surfaces are due to selective accretion of circumstellar (CS) material.
Due to gravitational settling and radiative acceleration, it is then mixed
in the shallow convection zone of the star. This explains why the anomalous abundance 
pattern is similar to that found in the gas phase of the interstellar medium (ISM) in which 
refractory elements like iron and silicon have condensed into dust grains. 

Later on, \citet{Kam02} and \citet{Mar09} developed a model which describes the interaction 
of the star with its local ISM and/or CS environment. As a result, different levels of underabundance 
are produced by different amounts of accreted material relative to the photospheric mass. 
The small fraction of this star group on the main-sequence (MS) is explained by the low probability 
of a star-cloud interaction and by the effects of meridional circulation, which dissolves any 
accretion pattern a few million years after the accretion has stopped. The hot end of this model is 
due to significant stellar winds for stars with $T_\mathrm{eff}$\,$>$\,12\,000\,K whereas 
the cool end, at about 6500\,K, is defined by convection which prevents the accreted 
material manifesting at the stellar surface.

Strong support for the selective accretion scenario has been given by \citet{Fol12} who found that 
half of their sample of Herbig Ae/Be stars exhibit the characteristic $\lambda$ Bootis type abundance 
pattern. We know that the density of CS material around Herbig Ae/Be stars is very high, perfectly
suited as the source for accretion.

Almost all $\lambda$ Bootis stars are located within the classical $\delta$ Scuti/$\gamma$ Doradus 
instability strip. \citet{Pau02a} presented a detailed analysis of their pulsational behaviour. 
They concluded that at least 70\% of the group members inside the classical instability strip pulsate, 
and they do so with high overtone p-modes (Q\,$<$\,0.020\,d). 

In this paper, we present the newly detected $\gamma$ Doradus pulsational characteristics of the classical
$\lambda$ Bootis star, HD 54272. Interesting enough, it was misidentified as a RR Lyrae type star by 
\citet{szczygiel2007}. We performed a detailed time series analysis of the Wide Angle Search for Planets
(WASP) and All Sky Automated Survey (ASAS) photometric data. Four statistically significant frequencies
and their combinations were detected. Our findings are discussed in comparison
with the first hybrid $\lambda$ Bootis/$\gamma$ Doradus
star, HR 8799. The latter is a very interesting young object hosting at least four planets and a 
massive dusty debris disk \citep{Esp13}. HD 54272, on the other hand, is a fast rotating star almost at the TAMS
with no detected IR excess. From spectroscopy, we deduce a metallicity of $-$0.8 to $-$1.1\,dex compared to
the Sun. This fact together with the high projected rotational velocities, make this object an interesting
and important test case for models dealing not only with $\gamma$ Doradus pulsation, but also with
selective accretion in the presence of meridional circulation.

\begin{table}
  \centering
  \begin{minipage}{55mm}
    \caption{The astrophysical parameters of HD 54272 \citep[][this work]{Pau02b} and HR 8799 \citep{Gra99,Bai12}. The errors 
		in the final digits of the corresponding quantity are given in parenthesis.}
    \label{stars_param}
    \centering
    \begin{tabular}{lcc}

      \hline\hline

      \multicolumn{1}{l}{\multirow{2}{*}{Quantity}} & HD 54272 & HR 8799           \\      
      & \multicolumn{2}{c}{values}   \\
      
  \hline
  
  \Teff [K]                      & 7010(217)         & 7430(75)    \\
  $[$M/H$]$                      & $-$0.8 to $-$1.1  & $-$0.47(10)  \\
  \logg                          & 3.83(10)          & 4.35(5)     \\
  $M_{\mathrm{V}}$ [mag]          & 2.33(30)          & 2.98(8)     \\
  $M$ [M$_{\sun}$]               & 1.69(19)          & 1.47(30)    \\
  $R$ [R$_{\sun}$]               & 2.2(3)            & 1.34(5)     \\
  $L$ [L$_{\sun}$]               & 1.01(12)          & 0.69(5)     \\
  \logt                         & 9.12              & $<$\,8.00   \\
  $v\,\sin\,i$ [km\,s$^{-1}$]    & 250(25)           & 38(2)       \\    

  \hline \\  

    \end{tabular}                                          
  \end{minipage}
\end{table}


%
\begin{figure}
\begin{center}
\includegraphics[width=85mm,clip]{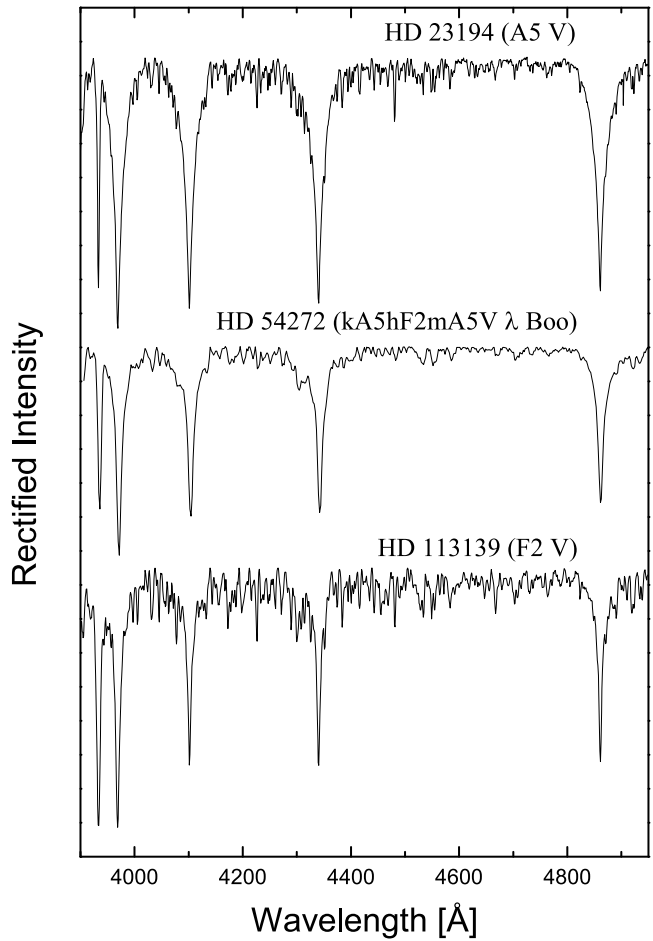}
\caption{The classification resolution spectrum of HD 54272 together with the two standard
stars HD 23194 (A5 V) and HD 113139 (F2 V). The latter were taken from \citet{Gra03} and have
the same resolution as our spectrum.}
\label{spec_hd54272} 
\end{center} 
\end{figure}

%
\begin{figure}
\begin{center}
\includegraphics[width=85mm,clip]{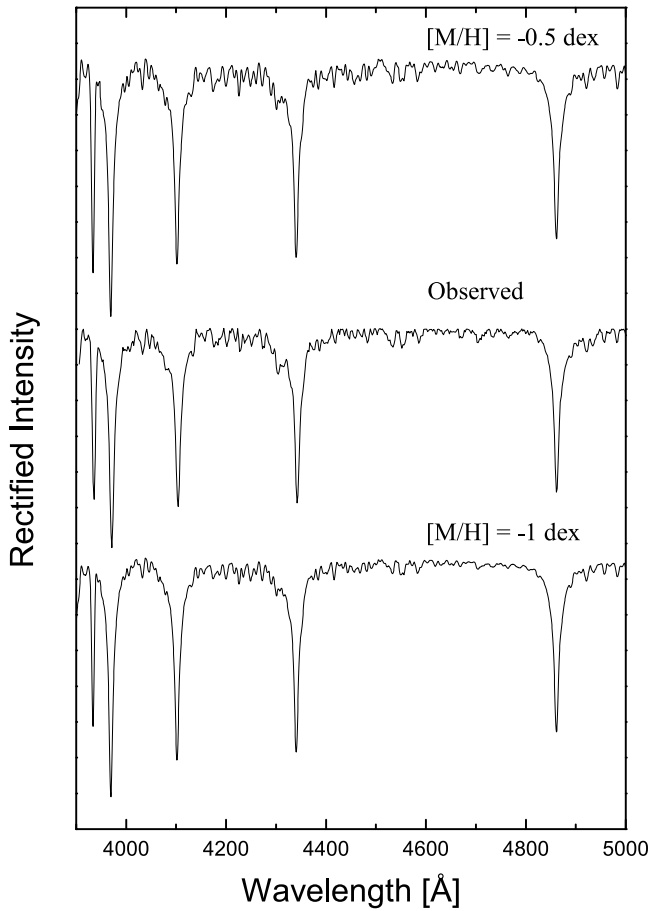}
\caption{The observed and synthetic (\Teff\,=\,7000\,K and \logg\,=\,3.8) spectra for HD 54272.}
\label{synthetic} 
\end{center} 
\end{figure}

%
\begin{figure}
\begin{center}
\includegraphics[width=85mm,clip]{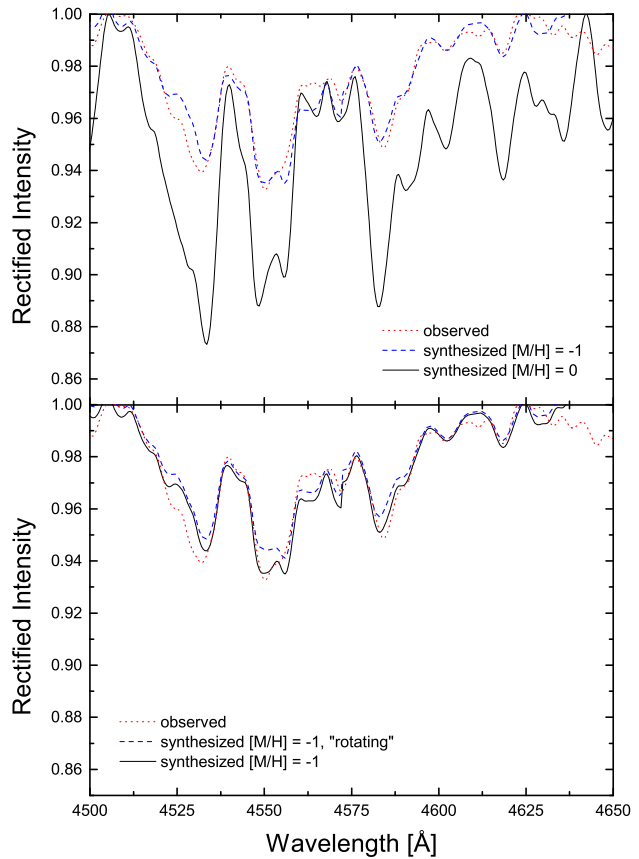}
\caption{The observed and two synthetic spectra showing the effect of metallicity (upper panel). The generated
`rotating' model (see text) is not very different to the non-rotating one (lower panel).}
\label{synthetic_two} 
\end{center} 
\end{figure}

%
\section{The characteristics of HD 54272 and HR 8799}

HD 54272 (BD+19 1620, V\,=\,8.80\,mag) was first classified as a member of the $\lambda$ Bootis group
by \citet{Pau01}. For this star, neither a Hipparcos parallax nor a radial velocity is available.
\citet{Pau02a} searched for $\delta$ Scuti type pulsations and found a null result with an upper limit
of 1.4\,mmag within a time series of 2.6 hours. It has to be emphasized that long term trends, such
as the here reported $\gamma$ Doradus pulsation, were not investigated. The only detailed follow-up analysis of the astrophysical parameters (Table \ref{stars_param}) for 
HD 54272 were presented by \citet{Pau02b}. The mass, effective temperature and surface gravity are
typical for an early F-type MS star. From isochrone fitting, they deduce that it has already spent
83\% of its MS life-time.

In this work we use the refined Yerkes classification (often denoted as `MK system') scheme 
introduced by \citet{Gra87}. It uses the notation `k', `h', and `m' for the classification of the Ca II K, hydrogen, 
and metal lines according to standard stars. It allows for a more detailed stellar classification than the
classical Yerkes system.

The spectrum of HD 54272 was observed at the Osservatorio Astronomico di Padova-Asiago with the 182\,cm telescope 
using the Boller \& Chivens spectrograph (600 lines mm$^{-1}$ grating) which gave a 
nominal resolution of 2.0\,\AA\,pixel$^{-1}$ (dispersion of 85\,\AA\,mm$^{-1}$) and a spectral coverage of about 1200\AA.

In Fig. \ref{spec_hd54272} the observed spectrum is shown together with two standards (HD 23194 and HD 113139). The latter were 
taken from \citet{Gra03} and have a resolution of 1.8\,\AA\,pixel$^{-1}$. Clearly, the high rotational velocity and the low 
metallicity of HD 54272 is visible. It is classified as `kA5hF2mA5V $\lambda$ Boo'.

In order to check the astrophysical parameters of HD 54272 and to search for an IR
excess, we made use of the spectral energy distribution (SED) fitting tool by \citet{Rob07}. 
As input data we used the available $UBV$, 2MASS and WISE photometry. The effective temperature and
surface gravity are well reproduced with the best fitting model. Furthermore, no IR excess was detected up to
22.1\,$\mu$m.

We next investigated the metallicity of HD 54272 in more detail. For this, 
a synthesized spectrum was computed using the program SPECTRUM\footnote{http://www.appstate.edu/$\sim$grayro/spectrum/spectrum.html} 
\citep{Gra94} and modified versions of the ATLAS9 code taken from the Vienna New Model Grid of Stellar Atmospheres, 
NEMO\footnote{http://www.univie.ac.at/nemo} \citep{Hei02}. We used a stellar atmosphere with the following parameters (Table \ref{stars_param}):
\Teff\,=\,7000\,K, \logg\,=\,3.8, and \vmic\,=\,2\,km\,s$^{-1}$. The synthetic spectrum was first folded with the instrumental
profile and then with different rotational profiles yielding a best fit for 250\,km\,s$^{-1}$ with an uncertainty of about 25\,km\,s$^{-1}$.
To test these parameters, a grid of atmospheres with effective temperatures and surface gravities around the input values were applied. The hydrogen
lines are best fitted with the original values with the constraint that they are not sensitive to \logg. To estimate the [M/H] value, we used 
different models from +0 to $-$2\,dex. Figure \ref{synthetic} shows the result for [M/H]\,=\,$-$0.5 and $-$1.0\,dex, respectively.
The median of the difference between the observed and synthetic spectrum with [M/H]\,=\,$-$1.0\,dex is only 0.2\%. Therefore we
adopted this values for HD 54272. For the estimation of its error, we used models with $-$0.5 and $-$1.5\,dex and calculated, again, the
differences of the synthetic and observed spectra. We notice that the difference with the more abundant model is smaller than the one
with the more underabundant one. From this, we conclude that the metallicity range is between $-$0.8 and $-$1.1\,dex. Figure
\ref{synthetic_two} shows the effect of the underabundant model for a typical metallic-line region around 4600\AA.

We checked if the high rotational velocity is affecting the derived metallicity. Because no reliable stellar atmospheres
taking into account the rotation, are available, coplanar ones are used. Assuming that the line-of-sight is almost
equator-on, the models of \citet{Sle80} are used to find the \Teff\ and \logg\ values at the poles. From their Table 1, the model
for a spectral type of A5 and a ratio of the angular velocity to the critical angular velocity of 0.9 almost
perfectly fits the parameters of HD~54272. It implies \Teff\,=\,9130\,K and \logg\,=\,4.25 at the poles. 
Again, we applied the above mentioned method to generate synthetic spectra using a different set of astrophysical
parameters each 15$\degr$ along the disk and assuming rigid body rotation. For the final spectrum, we integrated all individual spectra
over the stellar disk. Figure \ref{synthetic_two} shows the effect of the
rotation using stellar atmospheres with [M/H]\,=\,$-$1.0\,dex. Due to the high rotation velocity and the low abundance, the effect
is only small and still in the range of the derived uncertainty. This is line with the results of \citet{Tak08} investigated the effect of stellar rotation on the 
elemental abundance for A-type stars. They found no trends
for \vsini\ values up to 250\,km\,s$^{-1}$ (see Fig. 10 therein). As a conclusion, we are confident that the derived metal-weakness
is indeed intrinsic and not due to the inadequate treatment of stellar rotation.

\begin{figure}
\begin{center}
\includegraphics[width=75mm,clip]{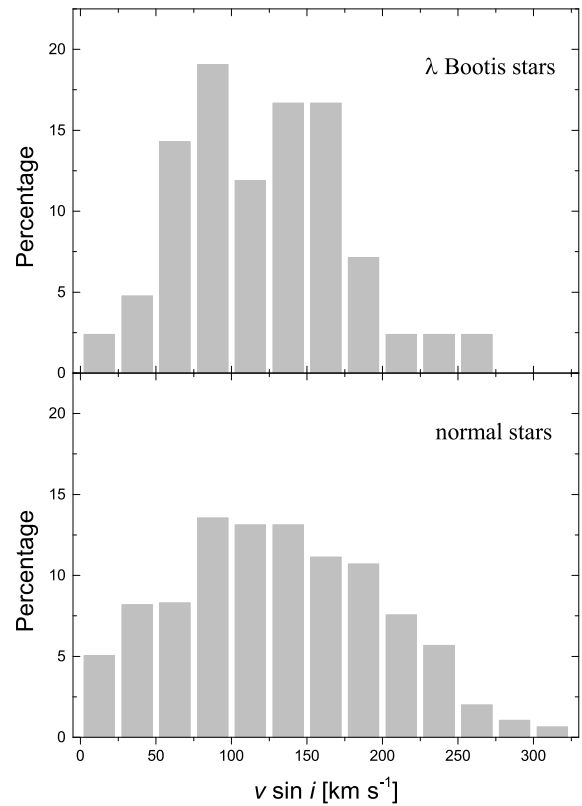}
\caption{The \ensuremath{{\upsilon}\sin i} distributions of $\lambda$ Bootis (upper panel)
and apparent `normal' type (lower panel) stars. According to a pair-sample t-test, both distributions
are on a 0.05 level, identical.}
\label{vsini} 
\end{center} 
\end{figure}

According to the formula that the rotation becomes imperceptible when
\ensuremath{{\upsilon}\sin i}\,$\leq$\,$D$\,=\,85\,\AA\,mm$^{-1}$,
we deduce a lower limit of 85\,km\,s$^{-1}$ for our spectrum. From our spectral synthesis, we derive a \ensuremath{{\upsilon}\sin i} of
about 250$\pm$25\,km\,s$^{-1}$. Using the mass and radius as listed in Table \ref{stars_param}, we calculate a 
break-up velocity for HD~54272 of 313\,km\,s$^{-1}$. Adopting an equatorial line-of-sight, this implies that HD 54272
rotates at least with 80\% of its break-up velocity. Such a high velocity is exceptional not only for $\gamma$ Doradus pulsators
\citep{Bru08}, but also for MS stars close to the TAMS, in general \citep{Zor12}. The astrophysical characteristics of HD~54272 fit
within the $\lambda$ Bootis group. In addition, we investigated the \ensuremath{{\upsilon}\sin i} distribution of the $\lambda$ Bootis stars, taken from
\citet{Pau02a}, and apparent `normal' type stars. The latter are from the list of \citet{Roy07} which covers the corresponding spectral range, 
excluding the late B-type stars and objects marked as `CB' (binaries) and `CP' (peculiar stars). The samples consist of 42 $\lambda$ 
Bootis and 953 `normal' stars, respectively. Figure \ref{vsini} shows the distributions of both groups. We performed a pair-sample t-test \citep{Ree87}
which results that both distributions are on a 0.05 level, identical. Therefore, \ensuremath{{\upsilon}\sin i} seems not to be an intrinsically `critical' 
astrophysical parameter for the formation and stability of the $\lambda$ Bootis phenomenon. {\bf However, the fast rotation of HD 54272 is not the common
characteristics for $\lambda$ Bootis stars (Fig. \ref{vsini}). But there is at least one similar member, HD~193256 (HIP~100286), with a \ensuremath{{\upsilon}\sin i} of
about 250\,km\,s$^{-1}$. It is also close to the TAMS but hotter (7700\,K).}

HR 8799 (HD 218396, V\,=\,5.95\,mag) is an exceptional, close-by (about 40\,pc), star classified as 
`kA5hF0mA5V $\lambda$ Boo' by \citet{Gra99}.
They derived the astrophysical parameters as listed in Table \ref{stars_param}. The typical $\lambda$ Bootis type
abundance pattern was later confirmed by \citet{Sad06}. The membership of this star to the Columba 
Association which has an age of about 30\,Myr is still a matter of debate \citep{Zuc11}. Moreover, \citet{Bai12}
summarizes the age determinations for this star (see Table 1 therein) which range from 30\,Myr to 1.6\,Gyr. The latter
has been deduced from an asteroseismic study by \citet{Moy10a}. We adopted the most probable one, $<$\,100\,Myr, as
argued and analysed by \citet{Bai12}.
HR 8799 is hosting at least four planets and a massive dusty debris disk \citep{Esp13}.
The $\gamma$ Doradus type pulsation of HR 8799 was studied by \citet{Zer99}. They found four 
significant frequencies (1.9791, 1.7268, 1.6498, and 0.2479\,d$^{-1}$) with corresponding amplitudes in Johnson $V$
between 15.98 and 5.66\,mmag. \citet{Moy10b} concluded from a detailed asteroseismic
study that HR 8799 is intrinsically metal-weak which is in contrary to the common accepted theory that the 
$\lambda$ Bootis phenomenon is restricted to the photosphere alone. However, one has to keep in mind that there
are at least two important free parameters (the inclination angle and the correct mode identification) which are vital 
for the asteroseismic analysis. Such an intrinsic metal-weakness, compared to the Sun, is still in the range which is found 
in the solar neighborhood \citep{Luc05}. If we accept this fact then also a non-solar isochrone has to be used to derive
the age of HR 8799. For a given temperature and luminosity, a lower metal isochrone would result in an older age. 
For the listed metallicity, it would naturally explain the often quoted discrepancy between the age of the Columba 
Association (30\,Myr) and some other determinations indicating an older age (100\,Myr). Furthermore, we have to ask
(i) if all `classical' $\lambda$ Bootis stars are intrinsic metal-weak or not, and (ii) if HR 8799 is a true member of the group.

\begin{table}
  \centering
  \begin{minipage}{70mm}
    \caption{Data characteristics and frequencies detected in the
    WASP and ASAS data with their identification. The uncertainties in the final digits are given in parenthesis.}
    \label{frek}
    \begin{tabular}{lllr}

      \hline\hline

      \multicolumn{1}{l}{\multirow{2}{*}{ID}}   &  Frequency  &	Amplitude & \multicolumn{1}{r}{\multirow{2}{*}{$S/N$}} \\

                                                &  [d$^{-1}$] & [mag]      &      \\ \hline
      
      \multicolumn{4}{c}{WASP} \\
      \multicolumn{2}{c}{Points available/used} & \multicolumn{2}{c}{Time base (used)} \\

      \multicolumn{2}{c}{19704/9949}						& \multicolumn{2}{c}{HJD 2454057-2455623}	\\								
      $f_{1}$ 									&	1.410117(2)	& 0.02128(7) 	& 16.0\\ 
      $f_{2}$ 									&	1.283986(5)	& 0.0138(1) 	&	6.3	\\
      $f_{3}$ 									&	1.293210(4)	& 0.01454(7)	&	6.8	\\	
      $f_{4}$  									&	1.536662(9)	& 0.00664(7)	&	5.5	\\
      $f_{5}$ 									&	1.15722(1)	& 0.00546(8)	&	5.9	\\
      $f_{6}$ 									&	0.22658(1)	& 0.00523(8)	&	6.4	\\
      \multicolumn{4}{c}{ASAS} \\
      \multicolumn{2}{c}{Points available/used} & \multicolumn{2}{c}{Time base (used)} \\
      \multicolumn{2}{c}{615/541}						& \multicolumn{2}{c}{HJD 2452621-2455166}	\\	
      $f_{1}$ 									&	1.41011(1)	& 0.0232(1) 	& 7.8 \\ 
      $f_{2}$ 									&	1.28399(2)	& 0.0147(1) 	&	6.7	\\
      $f_{3}$ 									&	1.293(6)		& 0.009(3)		&	4.8	\\
      & 1.89(3)			& 0.007(2)		& 4.4	\\
      & 5.01(2)			& 0.007(2)		& 4.0	\\
      
      \hline \\  
    \end{tabular}    
    \end{minipage}
\end{table}

%

\begin{figure}
\begin{center}
\includegraphics[width=75mm,clip]{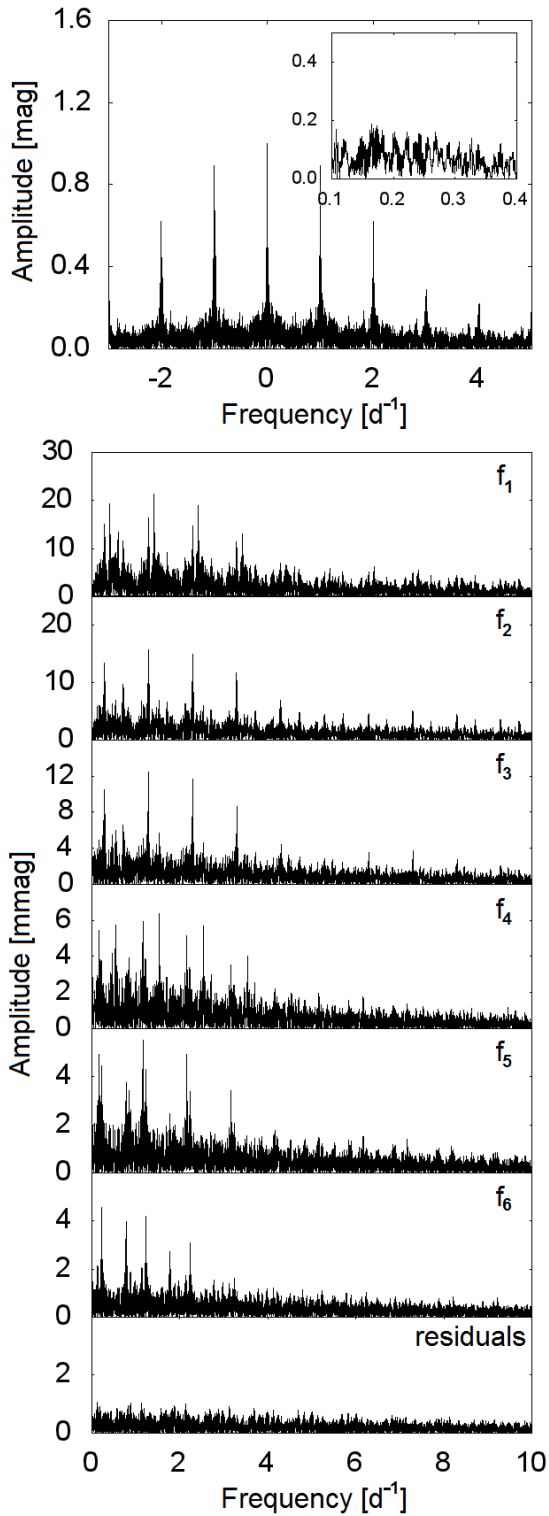}
\caption{The different steps of the frequency spectrum analysis for the WASP data. Top panel shows the spectral window dominated by daily aliases. 
The bottom part shows the highest peaks (identified in top left corner of each plot) during prewhitening. The lowest panel shows the residuals 
after prewhitening with frequencies $f_{1}$ to $f_{6}$.}
\label{fourier} 
\end{center} 
\end{figure}


%
\section{The pulsational behavior of HD 54272}

We analysed the photometric data from WASP \citep[][]{pollaco2006} and 
the All Sky Automated Survey \citep[ASAS\footnote{http://www.astrouw.edu.pl/asas/}, e.g. ][]{pojmanski1997,pojmanski2001} surveys. 
The WASP data, with the time base of 1566~d, 
are strongly affected by instrumental artifacts. Therefore, a Fourier model with six harmonics was fitted to the data 
and measurements deviating more than 2$\sigma$ were removed. This procedure was applied three times. For the ASAS data (time base of 2545~d), no such 
procedure was necessary, but the best quality data with flag `A' and `B' were used and few outliers were removed manually. 
Since ASAS gives information about the brightness in five different apertures, a weighted mean value was used.
A frequency analysis was performed 
using \textsc{Period04} \citep{lenz2004} software, 
which was also used for the determination of the uncertainties via Monte Carlo simulations.

ASAS data showed only five peaks: $f_{1}$, $f_{2}$, $f_{3}$ and two peaks with no clear interpretation. The latter are probably 
due to the intrinsic characteristics of the data set and the accuracy of the individual measurements. In the WASP data, six significant 
peaks were identified (see Table \ref{frek}). To estimate the reliability of these frequencies, the data were prewhitened with their various 
combinations. Frequencies, which were left out, were always detected again. In addition, we performed the complete time series analysis for 
two sub data sets. The overall data set is divided by the time of observations to HJD 2450000+ [5137:5186] and [5500:5624], respectively. 
This procedure should avoid the influence of seasonal effects on the Fourier analysis and the detection of possible spurious frequencies. 
Within the errors, all frequencies listed in Table \ref{frek} were found in both data sets.

From Fig. \ref{fourier} it is apparent that frequency spectra are strongly affected by daily aliases, which can be expected 
when analysing data from single-side observations. This makes the frequency identification somewhat ambiguous and 
also $f_{i}\pm1$~d$^{-1}$ should be considered as possible frequencies. Without any additional observations peaks can be hardly 
interpreted and it is only possible to guess potential scenarios. Nevertheless, peaks with the highest amplitudes correspond 
to the same frequencies in ASAS and WASP data. Therefore peaks that were identified (Table \ref{frek}) should be of higher reliability than their 1-day aliases.

Considering the frequencies in Table \ref{frek}, some interesting features appear. The peaks $f_{2}$, $f_{4}$ and $f_{5}$ are 
almost equally spaced with $f_{m}\approx 0.12647$~d$^{-1}$ (calculated as average of spacings between frequencies) with 
respect to $f_{1}$: $f_{4}-f_{1}=0.12655 \approx f_{m}$, $f_{1}-f_{2}=0.12613\approx f_{m}$ and $f_{1}-f_{5}=0.25290 \approx 2f_{m}$. 
Since the spectral window of the data set (Fig. \ref{fourier} detail in top panel) do not show any suspicious peak 
near 0.1265~d$^{-1}$, we consider this spacing as intrinsic to the pulsation of the star.

\begin{figure}
\begin{center}
\includegraphics[width=85mm,clip]{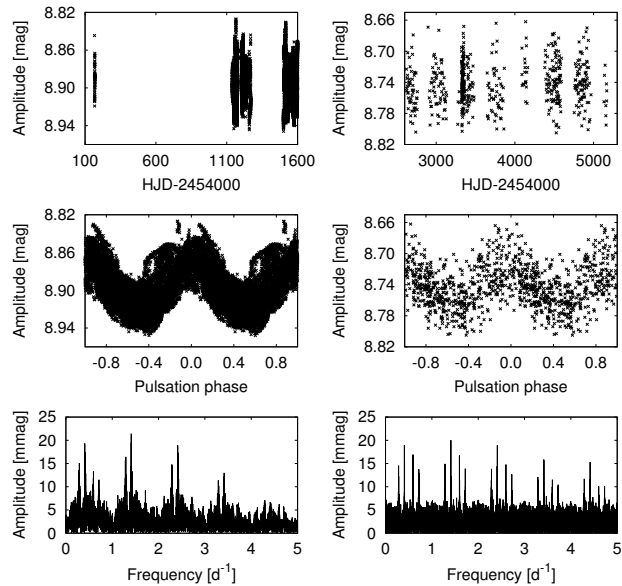}
\caption{The data distribution and phase plots of WASP (left panels) and ASAS data (right panels) 
folded with $f_{1}$ according to epoch HJD 2455164.116. The bottom panels show the frequency spectra of unprewhitened data.}
\label{phase} 
\end{center} 
\end{figure}

This symmetrical pattern is similar to those of modulated RR~Lyrae stars \citep{benko2011}, where the side-peaks are products 
of amplitude and/or phase modulation of the basic pulsation frequency with modulation period corresponding to the spacing between 
these peaks (in our case, $1/f_{m}=7.91$~d). Such an interpretation of the frequency spectra, probably led to the misclassification of HD 54272 
as a modulated RR Lyrae type star with a pulsation period of 0.7789192~d \citep[$1/f_{2}$ 
in our identification, ][]{szczygiel2007}. As Fig. \ref{phase} shows, the mean amplitude of light changes is only about 0.06~mag, 
which is too small for RR~Lyraes \citep{Smi03}. However, omitting small amplitude and two other frequencies, which were found, 
the light curve shape could easily be interpreted as a modulated RR~Lyrae star \citep{Ska13}.

We investigated if the found frequencies could be caused by rotation.
From the observational point of view, such rotational induced variability was never found for members of the
$\lambda$ Bootis group before \citep{Pau02a}. In general, as further ingredients, a local stellar magnetic field
which produces spots on the surface (as well as elemental overabundaces) and slow rotation in order to guarantee 
stability are needed \citep{Mes05}. This so-called  
Oblique Rotator model \citep{Sti50}, explains the variations as a geometrical effect as the star rotates, 
using a simple dipolar geometry. Because HD~54272 is a very fast rotator and magnetic fields were never detected for
$\lambda$ Bootis stars \citep{Boh90} it is very unlikely that the Oblique Rotator model could be applied. Based on 
the projected rotational velocity, radius (Table \ref{stars_param}), and $i$\,=\,0$\degr$ as well as the formula by \citet{Pre71},
we derive a rotational period of 0.45\,d for HD~54272. Larger values of the inclination would result in even shorter periods.
The fastest rotating objects with periods close to 0.5\,d are
very hot and high mass He-strong stars \citep{Mik10}. The $f_{i}+1$~d$^{-1}$ aliases of the first five frequencies match the 
expected rotational period. Also the regular spacing between frequencies could possibly be the consequence of rotation. However,
if we assume that the variability is due to rotation then we need a new model to explain this phenomenon because the Oblique
Rotator is, with our current knowledge, not applicable for HD~54272. Recently, \citet{Bal11a} investigated the light
curves of several thousands of A-type stars observed by the Kepler space mission. He found that almost 10\% of A-type stars exhibit variations 
resembling those usually attributed to starspots in cool stars, including a few exhibiting travelling waves usually interpreted as 
differentially rotating starspots. Most of these variations can also not be explained by the above mentioned Oblique Rotator model.
So, if we interpret the found variability of HD 54272 due to rotation, it would be the first $\lambda$ Bootis star with starspots
on its surface which would be a challenging example for models.

Explanation of peaks corresponding to $f_{4}=2f_{1}-f_{2}$ and $f_{5}=2f_{2}-f_{1}$ as coupling 
terms, which are often reported in $\gamma$ Doradus stars \citep{Bal11b} and which implies nonlinearity in 
the oscillation \citep[similar behavior was reported in HR 8799, ][]{Zer99}, fails, because $2f_{1}$ and $2f_{2}$ 
are not present in frequency spectra and none of daily aliases match the values needed for such consideration.

\begin{figure}
\begin{center}
\includegraphics[width=70mm,clip]{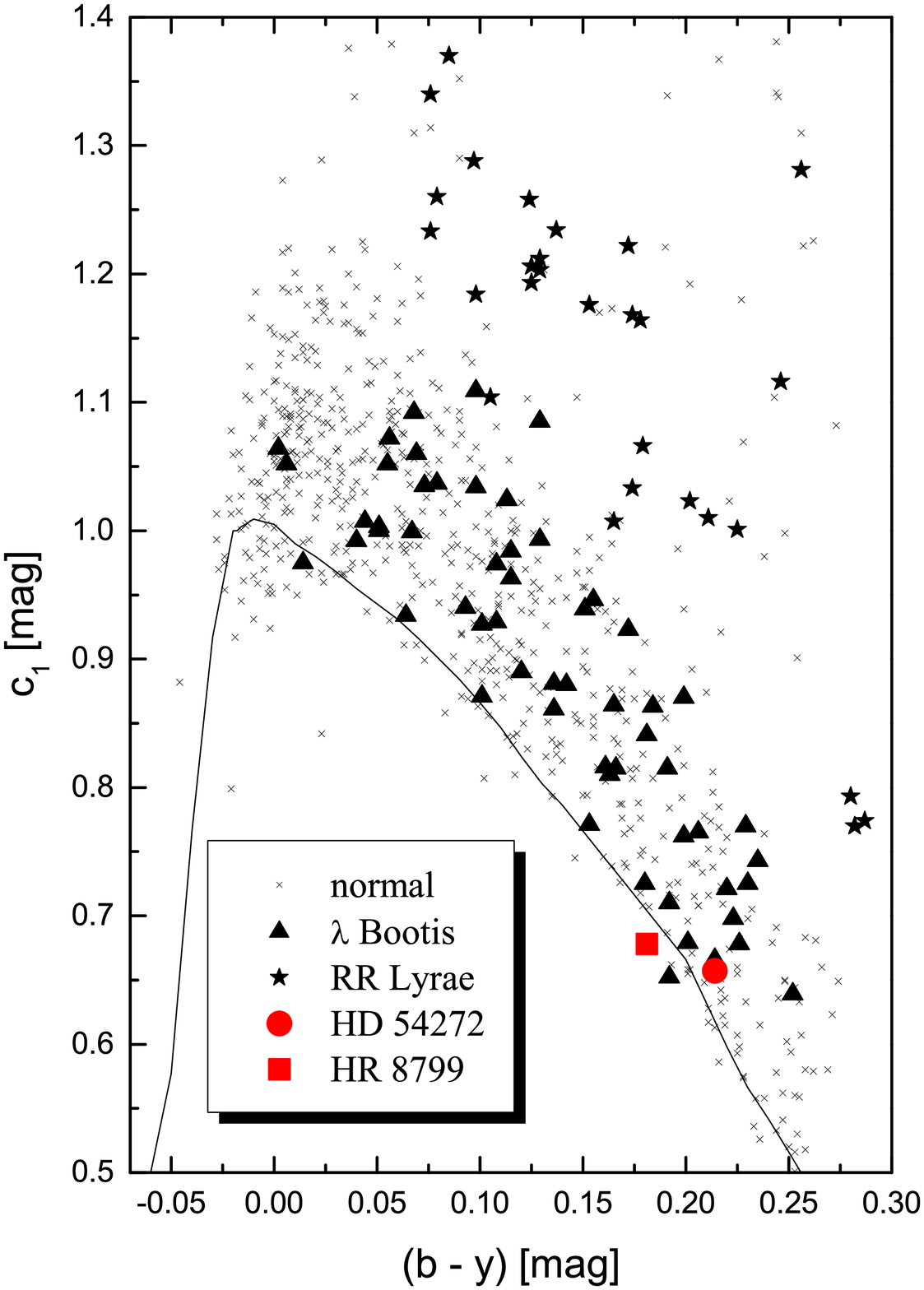}
\caption{The $(b-y)$ versus $c_{1}$ diagram for HD 54272, HR 8799, `normal' \citep[crosses, ][]{Gra87,Gra89a,Gra89b,Gar94}, 
$\lambda$ Bootis \citep[filled triangles, ][]{Pau02b}, and RR Lyrae stars \citep[open circles, ][]{Fer98}.
The solid line is the standard relation from \citet{Phi80}. The RR Lyrae stars are clearly separated from the $\lambda$ Bootis
stars and our target.}
\label{byc1} 
\end{center} 
\end{figure}

Furthermore, we checked the location of our target within the $(b-y)$ versus $c_{1}$ diagram which is sensitive 
to surface gravity \citep{Cra79}. For this purpose, we used data for `normal' \citep{Gra87,Gra89a,Gra89b,Gar94}, 
$\lambda$ Bootis \citep{Pau02b}, and RR Lyrae stars \citep{Fer98} together with the standard relation from \citet{Phi80}. 
The photometric indices were either taken from the listed references or from the catalogue by \citet{Hau98}. The
result is shown in Fig. \ref{byc1}. HD 54272 and HR 8799 are located where the other $\lambda$ Bootis stars
are situated. HR 8799 lies a little bit below the standard line which could be
due to its young age and the surrounding dense circumstellar material. The $\lambda$ Bootis stars are clearly separated from the RR Lyrae stars with two
exceptions: HD 148638 and HD 193281. Both stars are rather close visual binary systems with almost equally 
bright companions. Therefore, the available photometry could be influenced by the binarity.

As a conclusion, the presented analysis clearly shows that HD 54272 is not a RR Lyrae but a $\gamma$ Doradus pulsator
without a hint of $\delta$ Scuti variability \citep{Pau02a}.

%

%
\section{Conclusions}

We analysed the time series of the WASP and ASAS projects in order to shed
more light on the pulsational behaviour of HD 54272. This object was classified
as $\lambda$ Bootis star and RR Lyrae variable. A combination which, a-priori,
exclude each other. Our detailed analysis established this object as a $\gamma$ 
Doradus pulsator with six detected frequencies, which could be possibly biased by one day aliases. 
The amplitudes of detected peaks are between 5 and 21\,mmag. The three frequencies with the
highest amplitudes were detected in both data sets, independently. This example is
also important for other $\gamma$ Doradus stars which might be misidentified as RR Lyrae variables.

From spectroscopic data, we deduce a classification of `kA5hF2mA5V $\lambda$ Boo' for HD 54272.
A comparison with synthetic spectra yields a high rotational velocity (250$\pm$25\,km\,s$^{-1}$)
and a metal deficiency of $[$M/H$]$ of about $-$0.8 to $-$1.1\,dex compared to the Sun.
It is located very close to the TAMS.
These facts make HD 54272 an important test case for several different models including
pulsation, rotation, diffusion, and selective accretion.

Our results are compared with those from the first detected star which shows $\lambda$ Bootis
and $\gamma$ Doradus characteristics: HR 8799. Although the pulsational behaviours are very similar,
the evolutionary statuses are quite different. HR 8799 is a very young object hosting at least four 
planets and a massive dusty debris disk. If we accept that selective accretion plays a key role
for the $\lambda$ Bootis phenomenon, it seems to have a significant effect on $\gamma$ Doradus
type pulsation. This could be concluded from the fact that these two stars have widely different
evolutionary statuses and astrophysical parameters, but a strikingly similar pulsational characteristics.

The detection of further members of the $\lambda$ Bootis group showing $\gamma$ Doradus type pulsation
would put further constraints on models explaining and describing these phenomena. 

\section*{Acknowledgments}
The WASP Consortium consists of astronomers
primarily from Universities of St Andrews, Keele, Leicester,
Warwick, Queens University Belfast, The Open University, Isaac Newton Group La Palma
and Instituto de Astrofsica de Canarias. WASP-North is hosted
by the Issac Newton Group on La Palma and WASP-South is hosted by SAAO.
Funding for WASP comes from
consortium universities and from the UK Science and Technology Facilities
Council. 
This project is financed by the SoMoPro II programme (3SGA5916). The research leading
to these results has acquired a financial grant from the People Programme
(Marie Curie action) of the Seventh Framework Programme of EU according to the REA Grant
Agreement No. 291782. The research is further co-financed by the South-Moravian Region. 
It was also supported by the 
grant of Czech Ministry of Education, Youth and Sports 7AMB12AT003, 7AMB14AT015,
MUNI/A/0735/2012, and
the financial contributions of the Austrian Agency for International 
Cooperation in Education and Research (BG-03/2013 and CZ-10/2012).
We thank Markus Hareter and Martin Netopil for their assistance as well as the
referee for valuable comments.
This work reflects only the author's views and the European 
Union is not liable for any use that may be made of the information contained therein.

\label{lastpage}

\begin{thebibliography}{}
\bibitem[\protect\citeauthoryear{Baines et al.}{2012}]{Bai12} Baines E. K. et al., 2012, ApJ, 761, 57
\bibitem[\protect\citeauthoryear{Balona}{2011}]{Bal11a} Balona L. A., 2011, MNRAS, 415, 1691
\bibitem[\protect\citeauthoryear{Balona et al.}{2011}]{Bal11b} Balona L. A., Guzik J. A., Uytterhoeven K., Smith J. C., 
Tenenbaum P., Twicken J. D., 2011, MNRAS, 415, 3531
\bibitem[\protect\citeauthoryear{Benk{\H o} et al.}{2011}]{benko2011} Benk{\H o} J. M., Szab{\'o} R.,  Papar{\'o} M., 2011,
MNRAS, 417, 974
\bibitem[\protect\citeauthoryear{Bohlender \& Landstreet}{1990}]{Boh90} Bohlender, D. A., Landstreet, J. D., 1990, MNRAS, 247, 606
\bibitem[\protect\citeauthoryear{Bruntt, De Cat \& Aerts}{2008}]{Bru08} Bruntt H., De Cat P., Aerts C., 2008, A\&A, 478, 487
\bibitem[\protect\citeauthoryear{Crawford}{1979}]{Cra79} Crawford D. L., 1979, AJ, 84, 1858
\bibitem[\protect\citeauthoryear{Esposito et al.}{2013}]{Esp13} Esposito S. et al., 2013, A\&A, 549, A52
\bibitem[\protect\citeauthoryear{Fernley et al.}{1998}]{Fer98} Fernley J., Barnes T. G., Skillen I., Hawley S. L., Hanley C. J., 
Evans D. W., Solano E., Garrido R., 1998, A\&A, 330, 515
\bibitem[\protect\citeauthoryear{Folsom et al.}{2012}]{Fol12} Folsom C. P., Bagnulo S., Wade G. A., Alecian E., Landstreet J. D., 
Marsden S. C., Waite I. A., 2012, MNRAS, 422, 2072
\bibitem[\protect\citeauthoryear{Garrison \& Gray}{1994}]{Gar94} Garrison R. F., Gray R .O., 1994, AJ, 107, 1556 
\bibitem[\protect\citeauthoryear{Gray \& Garrison}{1987}]{Gra87} Gray R. O., Garrison R. F., 1987, ApJS, 65, 581
\bibitem[\protect\citeauthoryear{Gray \& Garrison}{1989a}]{Gra89a} Gray R. O., Garrison R. F., 1989a, ApJS, 69, 301 
\bibitem[\protect\citeauthoryear{Gray \& Garrison}{1989b}]{Gra89b} Gray R. O., Garrison R. F., 1989b, ApJS, 70, 623 
\bibitem[\protect\citeauthoryear{Gray \& Corbally}{1994}]{Gra94} Gray R. O., Corbally C. J., 1994, AJ, 107, 742
\bibitem[\protect\citeauthoryear{Gray \& Kaye}{1999}]{Gra99} Gray R. O., Kaye A. B., 1999, AJ, 118, 2993
\bibitem[\protect\citeauthoryear{Gray et al.}{2003}]{Gra03} Gray R. O., Corbally C. J., Garrison R. F., McFadden M. T., 
Robinson P. E., 2003, AJ, 126, 2048
\bibitem[\protect\citeauthoryear{Hauck \& Mermilliod}{1998}]{Hau98} Hauck B., Mermilliod M., 1998, A\&AS, 129, 431 
\bibitem[\protect\citeauthoryear{Heiter et al.}{2002}]{Hei02} Heiter U. et al., 2002, A\&A 392, 619 
\bibitem[\protect\citeauthoryear{Kamp \& Paunzen}{2002}]{Kam02} Kamp I., Paunzen E., 2002, MNRAS, 335, L45
\bibitem[\protect\citeauthoryear{Lenz \& Breger}{2004}]{lenz2004} Lenz P., Breger M., 2004, Comm. Ast., 146, 53
\bibitem[\protect\citeauthoryear{Luck \& Heiter}{2005}]{Luc05} Luck R. E., Heiter U., 2005, AJ, 129, 1063
\bibitem[\protect\citeauthoryear{Martinez-Galarza et al.}{2009}]{Mar09} Martinez-Galarza J. R., Kamp I., Su K. Y. L., G{\'a}sp{\'a}r A., 
Rieke G., Mamajek E. E., 2009, AJ, 694, 165
\bibitem[\protect\citeauthoryear{Mestel \& Moss}{2005}]{Mes05} Mestel L., Moss D., 2005, MNRAS, 361, 595
\bibitem[\protect\citeauthoryear{Michaud \& Charland}{1986}]{Mic86} Michaud G., Charland Y., 1986, ApJ, 311, 326
\bibitem[\protect\citeauthoryear{Mikul\'a\v{s}ek et al.}{2010}]{Mik10} Mikul\'a\v{s}ek Z., Krti{\v c}ka J., Henry G. W., de Villiers S. N., 
Paunzen E., Zejda, M. 2010, A\&A, 511, L7
\bibitem[\protect\citeauthoryear{Moya et al.}{2010a}]{Moy10a} Moya A., Amado P. J., Barrado D., Garc{\'i}a Hern{\'a}ndez A. G., Aberasturi M.,
Montesinos B., Aceituno F.,, 2010a, MNRAS, 405, L81
\bibitem[\protect\citeauthoryear{Moya et al.}{2010b}]{Moy10b} Moya A., Amado P. J., Barrado D., Garc{\'i}a Hern{\'a}ndez A. G., Aberasturi M.,
Montesinos B., Aceituno F., 2010b, MNRAS, 406, 566
\bibitem[\protect\citeauthoryear{Paunzen}{2001}]{Pau01} Paunzen E., 2001, A\&A, 373, 633
\bibitem[\protect\citeauthoryear{Paunzen et al.}{2002a}]{Pau02a} Paunzen E. et al., 2002a, A\&A, 392, 515
\bibitem[\protect\citeauthoryear{Paunzen et al.}{2002b}]{Pau02b} Paunzen E., Iliev I. Kh., Kamp I., Barzova I. S., 2002b, MNRAS, 336, 1030
\bibitem[\protect\citeauthoryear{Philip \& Egret}{1980}]{Phi80} Philip A. G., Egret D., 1980, A\&AS, 40, 199
\bibitem[\protect\citeauthoryear{Pojmanski}{1997}]{pojmanski1997} Pojma{\'n}ski G., 1997, AcA, 47, 467
\bibitem[\protect\citeauthoryear{Pojmanski}{2001}]{pojmanski2001} Pojma{\'n}ski G., 2001, IAU Colloq.~183: Small Telescope Astronomy on Global Scales, 246, 53 
\bibitem[\protect\citeauthoryear{Pollacco et al.}{2006}]{pollaco2006} Pollacco D. L. et al., 2006, PASP, 118, 1407 
\bibitem[\protect\citeauthoryear{Preston}{1971}]{Pre71} Preston G. W., 1971, PASP, 83, 571
\bibitem[\protect\citeauthoryear{Rees}{1987}]{Ree87} Rees D. G., 1987, Foundations of Statistics, Chapman \& Hall, London
\bibitem[\protect\citeauthoryear{Robitaille et al.}{2007}]{Rob07} Robitaille T. P., Whitney B. A., Indebetouw R., Wood K., 2007, ApJS, 169, 328
\bibitem[\protect\citeauthoryear{Royer et al.}{2007}]{Roy07} Royer F., Zorec J., G{\'o}mez A. E. 2007, A\&A, 463, 671
\bibitem[\protect\citeauthoryear{Sadakane}{2006}]{Sad06} Sadakane K., 2006, PASJ, 58, 1023
\bibitem[\protect\citeauthoryear{Skarka}{2013}]{Ska13} Skarka M., 2013, A\&A, 549, A101
\bibitem[\protect\citeauthoryear{Slettebak et al.}{1980}]{Sle80} Slettebak A., Kuzma T. J., Collins G. W., II, 1980, ApJ, 224, 171
\bibitem[\protect\citeauthoryear{Smith}{2003}]{Smi03} Smith H. A., 2003, RR Lyrae Stars, Cambridge University Press, Cambridge
\bibitem[\protect\citeauthoryear{Stibbs}{1950}]{Sti50} Stibbs, D. W. N., 1950, MNRAS, 110, 395
\bibitem[\protect\citeauthoryear{Szczygie{\l} \& Fabrycky}{2007}]{szczygiel2007} Szczygie{\l} D. M., Fabrycky D. C., 2007, MNRAS, 377, 1263
\bibitem[\protect\citeauthoryear{Takeda et al.}{2008}]{Tak08} Takeda Y., Han I., Kang D.-I., Lee B.-C., Kim K.-M., 2008, Journal of Korean Astronomical Society, 41, 83
\bibitem[\protect\citeauthoryear{Zerbi et al.}{1999}]{Zer99} Zerbi F. M. et al., 1999, MNRAS, 303, 275
\bibitem[\protect\citeauthoryear{Zorec \& Royer}{2012}]{Zor12} Zorec J., Royer F., 2012, A\&A, 537, A120
\bibitem[\protect\citeauthoryear{Zuckerman et al.}{2011}]{Zuc11} Zuckerman B., Rhee J. H., Song I., Bessell M. S., 2011, ApJ, 732, 61
\end{thebibliography}
\end{document}